\documentstyle[sprocl,epsfig]{article}

\def\be{\begin{equation}}
\def\ee{\end{equation}}
\def\bea{\begin{eqnarray}}
\def\eea{\end{eqnarray}}
\def\Hz{\rm\, Hz}

\begin{document}

\title{ADAPTIVE IDENTIFICATION OF {\sl VIRGO}-LIKE NOISE SPECTRUM}

\author{ E. CUOCO, G. CURCI }

\address{Department of Physics of Pisa University, P.zza Torricelli 2,\\ 
56100 - Pisa, Italy\\
E-mail cuoco@hpth1.difi.unipi.it\, curci@ipifidpt.difi.unipi.it}

\author{ M. BECCARIA }

\address{Department of Physics of Lecce University, Via Arnesano, \\
73100 - Lecce, Italy\\
E-mail beccaria@riscle7.le.infn.it}

\maketitle\abstracts{
The aim of this work is to show how it is possible to build 
an {\it on line} whitening filter in an adaptive way. 
We have modeled the {\sl VIRGO} spectrum as an autoregressive stochastic process, 
after a pre-filtering of the theoretical curve which flattens the low frequency part of the spectrum.  
We have tested some very popular adaptive algorithms, based
on the gradient methods and on the least squares methods with a lattice 
structure filter. }
\section{Modeling the {\sl VIRGO} noise spectrum}
The {\sl VIRGO}  noise spectrum is characterized by a wide band part and
by several spectral peaks, so it shows a lot of features, which we need to
identify very accurately to obtain a whitened spectrum to be used in data analysis algorithms and to keep under control a possible slow non stationarity 
of the noise.

Our theoretical curve for the Virgo-like spectrum contains:
the shot noise, the pendulum thermal noise, 
mirrors and violin modes:

\be
S(f)=
\frac{\displaystyle S_1}{\displaystyle f^5} + 
\frac{\displaystyle S_2}{\displaystyle f} + S_3\left( 1 + \left(\frac{f}{f_K}\right)^2\right) + S_v(f)
\ee
where
\bea
&f_K = 500 {\rm Hz}\qquad &S_1 = 1.08\cdot 10^{-36} \\
&S_2 = 0.33 \cdot 10^{-42} \qquad &S_3 = 3.24\cdot 10^{-46}
\eea
The contribute of violin resonances is given by 
\be
S_v(f) = \sum_n\frac{1}{n^4}\frac{f_1^{(c)}}{f} \frac{C_c \phi_n^2}
{\left(\displaystyle
\frac{1}{n^2}\frac{f^2}{f_1^{(c)2}} -1 
\right)^2+\phi_n^2} + (c\leftrightarrow f)
\ee
where we take into account the different masses of close and far mirrors,
being
\be
f_n^{(c)} = n \cdot 327{\Hz} \qquad
f_n^{(f)} = n \cdot 308.6\Hz 
\ee
\be
C_c  = 3.22\cdot 10^{-40}\qquad  C_f = 2.82\cdot 10^{-40} \qquad
\phi_n^2 = 10^{-7}
\ee
\vskip -1truecm
\begin{figure}[htb]
\begin{center}
\epsfig{file=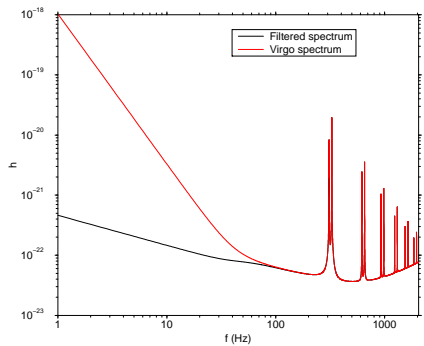,width=0.45\textwidth,angle=270}
\end{center}
\caption{Virgo-like noise spectrum and the result 
after a pre-filtering which flattens the low frequency tail. Sampling frequency $f_s=4096\Hz$.}
\label{fig:spettro}
\end{figure}

We need a pre-filtering of the low-frequency part of the
spectrum, because the pendulum mode dominates the autocorrelation
function. If we used adaptive algorithms to find the parameters 
of our spectrum model in such a way to follow the slow non stationarity
of the noise, we would need a short learning time for the algorithms. 
This is an impossible task if we analyze a noise characterized by a long
autocorrelation time. 
\subsection{Noise modeling as an AR process}
We fit the spectrum  showed in figure~\ref{fig:spettro} by an
autoregressive stochastic process of order $P$:
\be
x_n = \sum_{k=1}^P a_k x_{n-k} + \sigma\xi_n
\ee
where $\xi_n$ are independent normal random numbers and $a_k$ are the $P$ 
parameters of the model.

The relationship between the parameters of the model and the autocorrelation function $r_{xx}(n)$ is given by the {\it Yule--Walker} equations  
\be
r_{xx}(n)= \left \{
\begin{array}{ll}
\sum_{k=1}^p a_k r_{xx}(n-k) & \mbox{for}\  n\ge 1 \\
\sum_{k=1}^p a_k r_{xx}(-k)+\sigma^2 & \mbox{for}\ n=0\, 
\end{array}
\right.
\ee
Given the spectrum, we obtain the autocorrelation function and we can solve for the coefficients $\{a_k\}$ by the Durbin algorithm~\cite{Alex2}.

The key point is to find the optimal value $P$ for the order of the
process. In literature there are some standard criteria which may be used in order to determine this value. We have tested the Akaike information criterion (AIC), the Minimum description length (MDL) and the Akaike final prediction error (FPE): the MDL criterion is the most efficient among them
in finding a minimum. It reaches a minimum value for $ P=292$. 

\begin{figure}[htb]
\hbox{
\epsfig{file=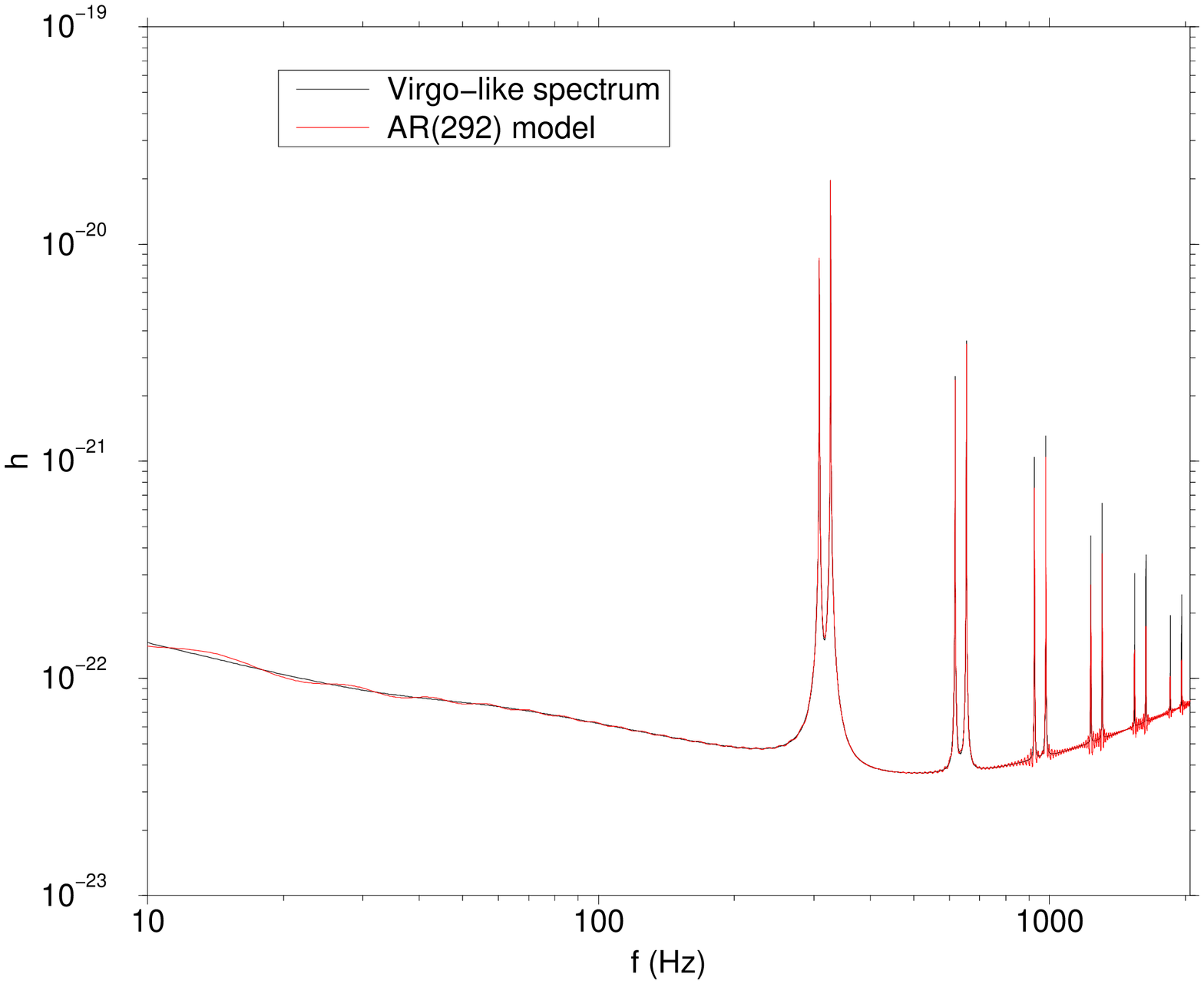,width=0.45\textwidth,angle=270}
\kern -0.8truecm
\epsfig{file=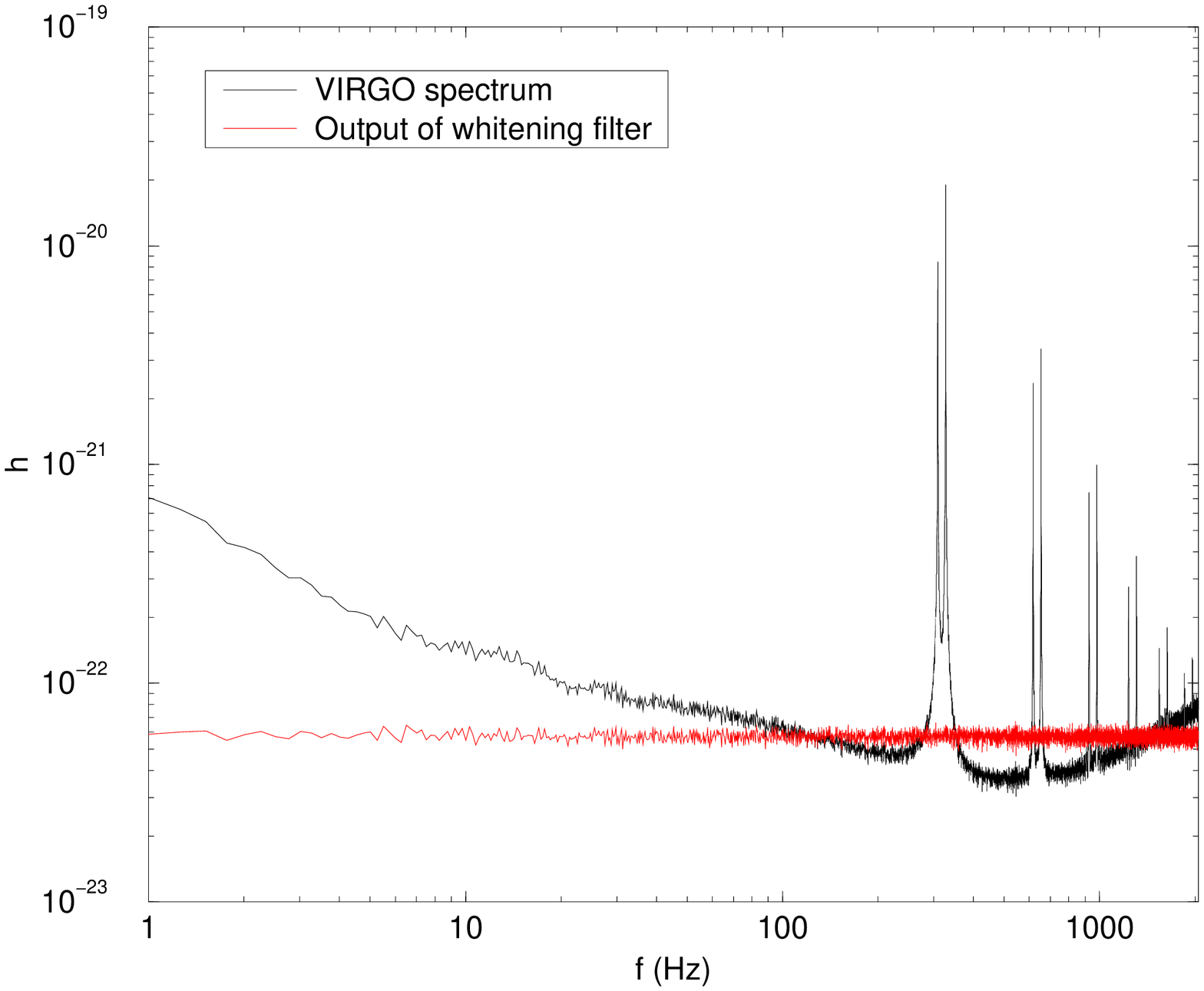,width=0.45\textwidth,angle=270}
}
\caption{On the left: AR(292) model fit to the spectrum. On the right: input and output of a Durbin whitening filter}
\label{fig:durbinwhite}
\end{figure}

We choose to fit the filtered noise spectrum with an $AR(292)$ model
and to generate the data sample on which we shall perform the adaptive test
with the AR parameters estimates with the Durbin algorithm. 
The result of the fit is shown in figure~\ref{fig:durbinwhite}.
Once we found the reflection parameters~\cite{Alex2} 
with Durbin algorithm we can implement the associated linear predictor 
in the time domain and in the lattice fashion~\cite{Alex2}.
The output of this whitening filter in shown in figure ~\ref{fig:durbinwhite}.
\vskip -0.7truecm

\section{Adaptive identification of AR parameters}
We have just seen how to implement a whitening filter if  we use
the Durbin algorithm to compute  the reflection coefficients of our process. 
Now, we want to estimate these coefficients {\it on line} without
estimating the correlation function first, but directly from the input data.
There are several ways of accomplishing this purpose. 
We have tested some of the most popular algorithms, which can be
divided in two main categories: the gradient methods (GAL) and 
the least squares methods (LS)
In the gradient based methods we use an estimate of the cost function
at the $n$th step which is based on the $n$th data input, while the updating
criterion for the learning parameter is derived by minimizing the 
{\it expectation} value of the cost function. 
On the other hand, in the least squares based methods the optimal least squares
prediction is computed at every point in time keeping into account the whole data history. 

\subsection{Results}
\begin{figure}[htb]
\hbox{
\epsfig{file=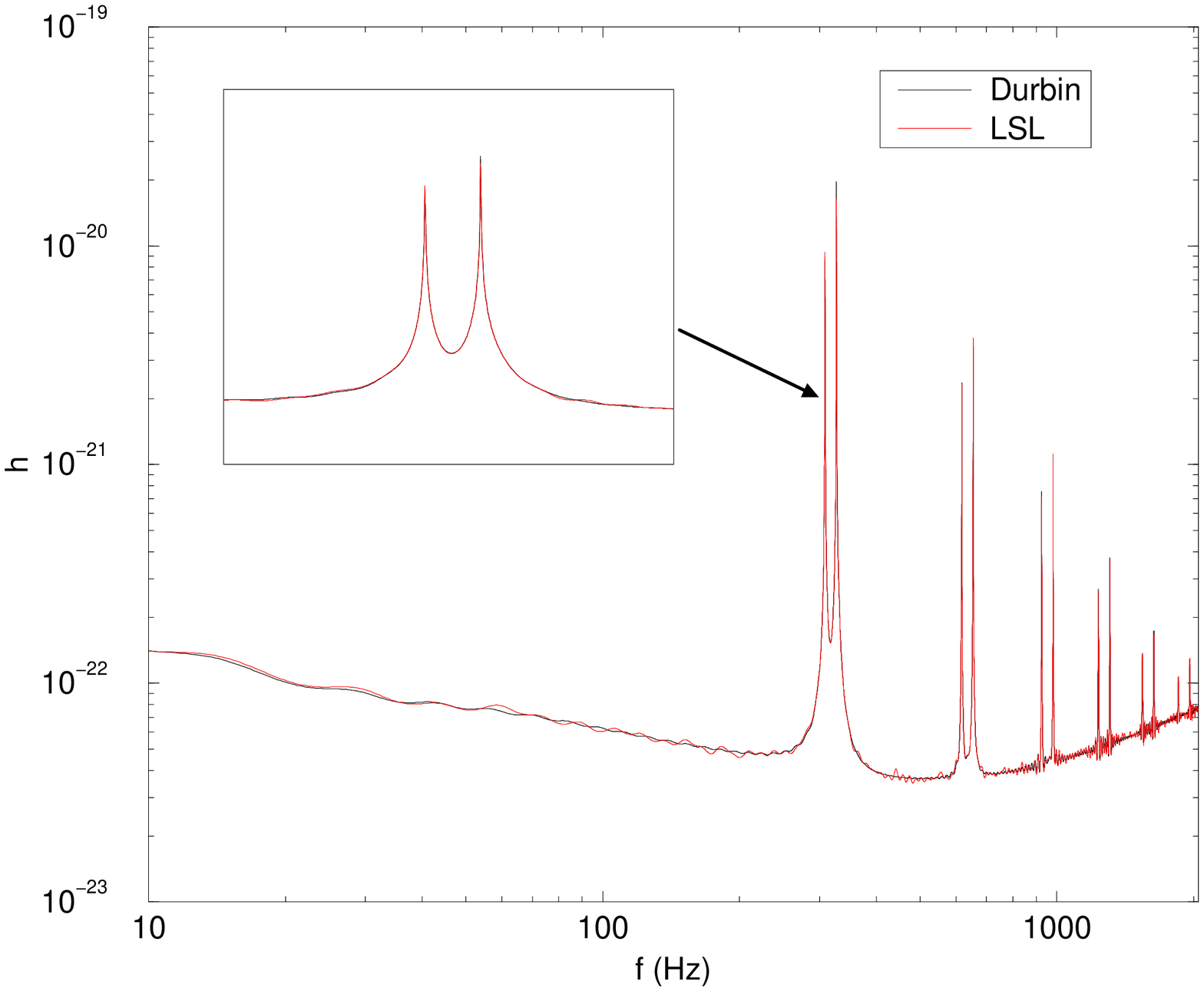,width=0.45\textwidth,angle=270}
\kern -0.8truecm
\epsfig{file=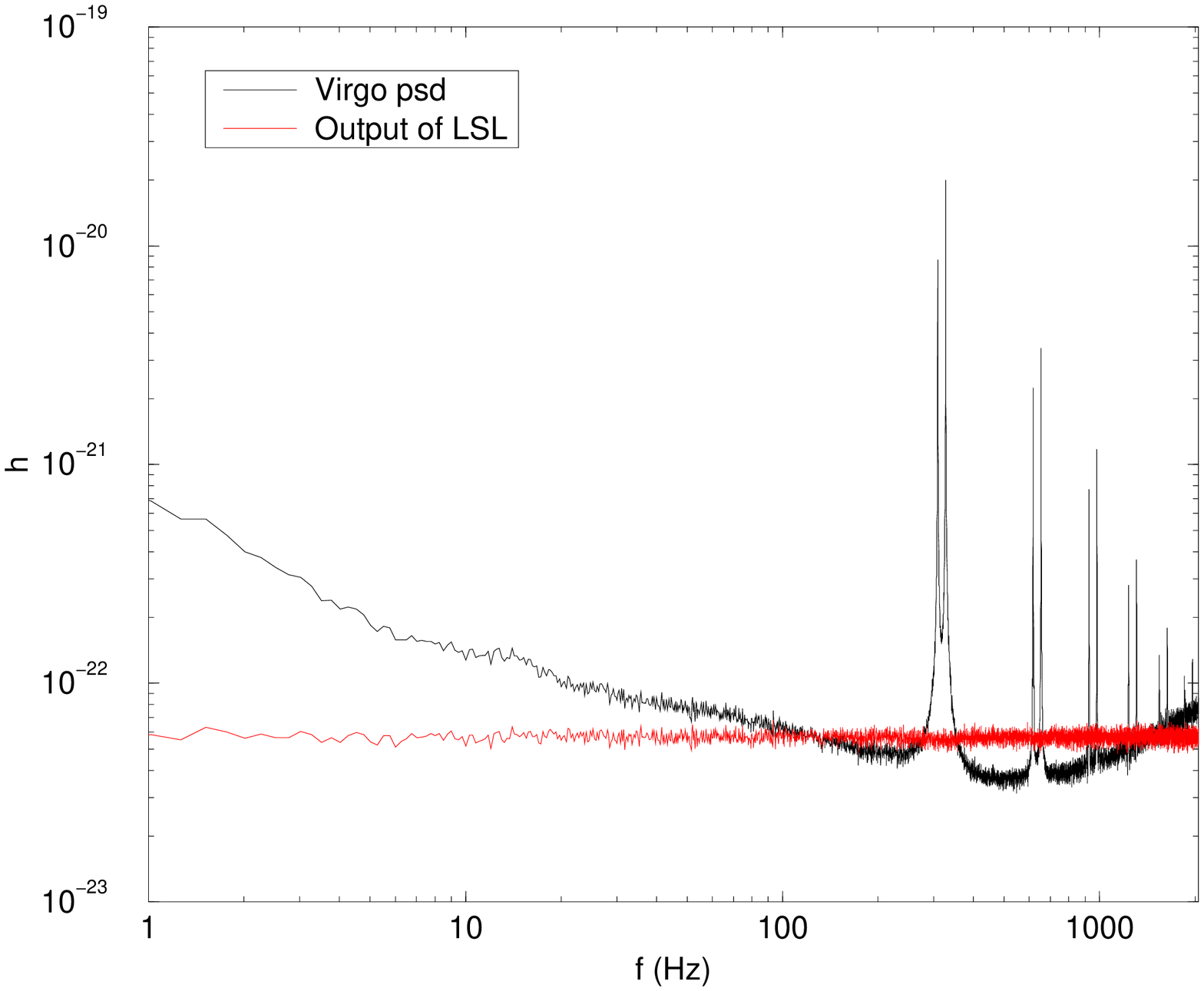,width=0.45\textwidth,angle=270}
}
\caption{LSL fit to AR(292) spectrum. Input and output of a LSL filter.}
\label{fig:lsl}
\end{figure}

We report in figure~\ref{fig:lsl} the results obtained with the Least Squares Lattice (LSL) 
filter which is the best working among the algorithms we analyzed. 
It is evident the efficiency of this algorithm in following all the features 
of the noise spectrum. 
\vskip -1truecm

The fast convergence of the algorithm lets us follow non stationarity
of the noise which are slower than one minute. 
To check the quality of an estimator we need to put a bound on its 
performance.
We can use general results from the theory of statistical estimators.
The variance of any unbiased estimator is bounded from below by the
Cramer-Rao bound.
 We checked if the variance of the AR parameters estimated with the LSL 
estimator attains these limits or not. 
\begin{figure}[!htb]
\hbox{
\epsfig{file=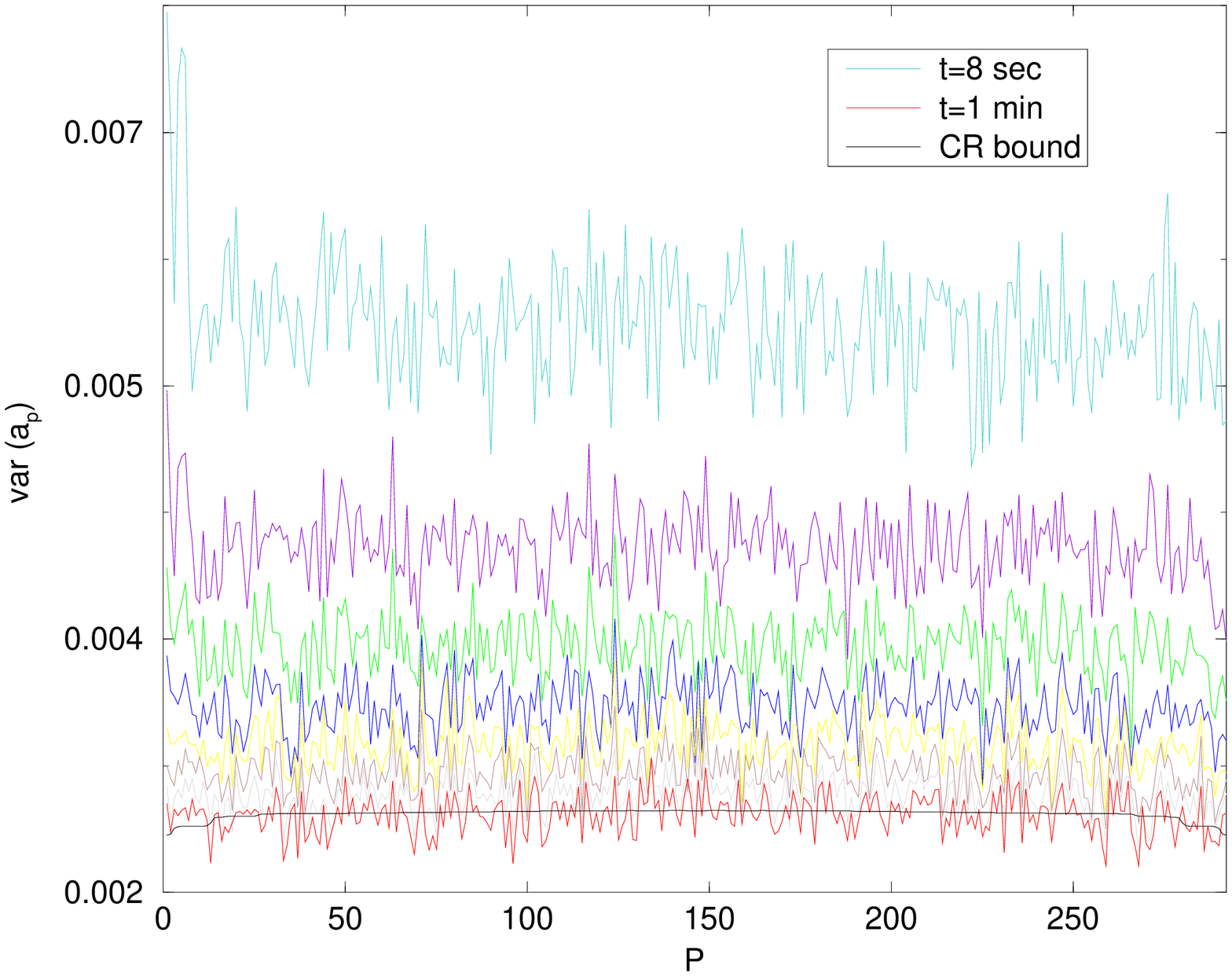,width=0.45\textwidth,angle=270}
\kern -0.8truecm
\epsfig{file=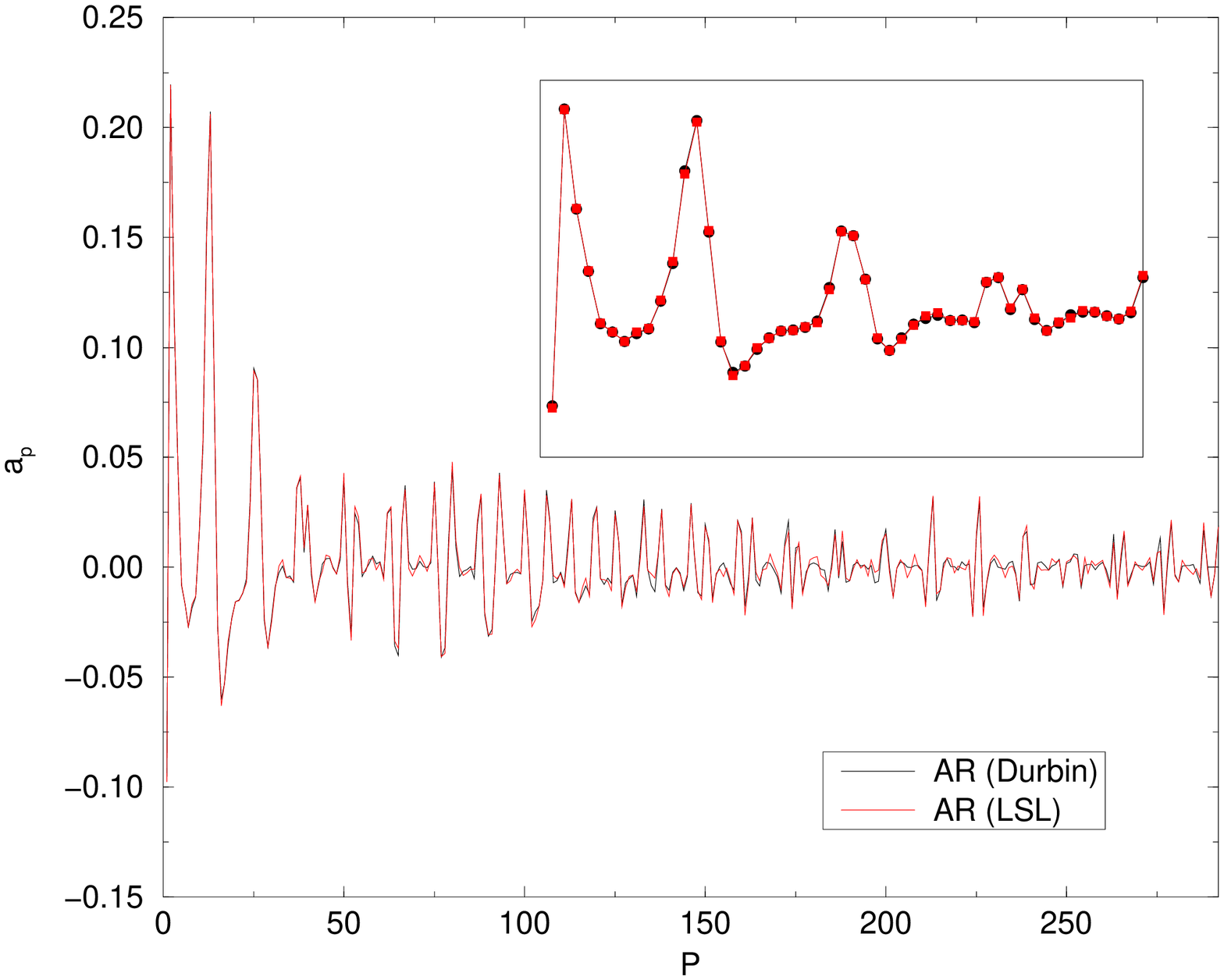,width=0.45\textwidth,angle=270} 
}
\caption{Cramer Rao lower bound. LSL fit to the AR parameters}
\label{fig:CRbound}
\end{figure} 
We have computed the variance of each parameter 
and of $\sigma^2$ at different times
 to follow how the variance approaches  the CR bound as time 
goes by.
The times are delayed each other by $8$ seconds, the last time corresponding 
to one elapsed minute. 

In Figure~\ref{fig:CRbound} we report the results. It is evident how
the variance of each coefficient flattens to the CR limit as the
number of iterations of the adaptive algorithm increases. 
After one minute of data input, the variance of the
coefficients has already reached the CR bound.
Therefore we conclude that the LSL estimator is an efficient one.
\section{Conclusions}
We have shown that it is possible to model a noise spectrum with complex 
features like those relevant for the VIRGO experiment by parameterizing it 
in terms of a small number of parameters.

We have tested some adaptive algorithms which are able to fit on line the parameters of an autoregressive representation of the 
{\sl VIRGO}-like spectrum. 
The most efficient of them is the least squares lattice 
algorithm which, after one minute of data, converges and reproduces all 
the desired spectral features attaining moreover the Cramer Rao lower bound.
 
In principle, the fast convergence lets us follow the slow non 
stationarity of the noise. 
In a forthcoming note we shall describe in a detailed way the 
efficiency of the algorithm in dealing with non stationarities
as a function of their characteristic time scales and amplitude magnitude.


\section*{References}
\begin{thebibliography}{99}

\bibitem{Alex2}
S.~Thomas Alexander, {\em Adaptive signal processing}, Springer, New York,
  1986.

\bibitem{Kay}
S.~M. Kay, {\em Modern spectral estimation}, Prentice-Hall, Englewood Cliffs,
  New Jersey, 1995.

\bibitem{Therrien}
Charles~W. Therrien, {\em Discrete random signals and statistical signal
  processing}, Prentice-Hall, Englewood Cliffs, 1992.

\end{thebibliography}
\end{document}